\documentclass[secnumarabic,amssymb, nobibnotes, aps, prd]{revtex4}%
\usepackage{graphicx}
\usepackage{dcolumn}
\usepackage{bm}
\usepackage{amsmath}%
\usepackage{amsfonts}%
\usepackage{amssymb}
\begin{document}

\title{Search for anisotropic light propagation as a function of laser beam alignment relative to the Earth's velocity vector}

\author{C. E. Navia\footnote{e-mail:navia@if.uff.br}, C. R. A. Augusto, D. F. Franceschine, M. B. Robba and K. H. Tsui}
\address{Instituto de F\'{\i}%
sica Universidade Federal Fluminense, 24210-346,
Niter\'{o}i, RJ, Brazil}

\date{\today}

\begin{abstract}
A laser diffraction experiment was conducted to study light propagation in air. The experiment is easy to reproduce and it is based on simple optical principles. Two optical sensors (segmented photo-diodes) are
used for measuring the position of diffracted light spots with a precision better than  $0.1\;\mu m$. 
The goal is to look for signals of anisotropic light propagation as function of the laser beam alignment to the Earth's motion (solar barycenter motion) obtained by COBE. Two raster search techniques have been used. First, a fixed laser beam in the laboratory frame that scans due to Earth's rotation. Second, an active rotation of the laser beam on a turntable system. The results obtained with both methods show that the course of the light rays are affected by the motion of the Earth, and a predominant quantity of first order with a $\Delta c/c=-\beta (1+2a)\cos \theta$ signature with $a=-0.4106\pm 0.0225$ describes well the experimental results. This result differs in a amount of 18\% from the Special Relativity Theory prediction and that supplies the value of $a=-1/2$ (isotropy).

\end{abstract}


\maketitle

\section{Introduction}

There are several physical reasons, theoretical and experimental, that could justify a search for anisotropies in light propagation. It is well known that Lorentz and Poincar$\acute{e}$  were the first ones to build the major part of the relativity theory on the basis of the ether concept as an inertial rest frame, and it is fully compatible with the Einstein's Special Relativity Theory (SRT). There are also some test theories of SRT, where the Lorentz transformations are modified. For example, an ether theory that maintains the absolute simultaneity and is kinematically equivalent to Einstein SRT was constructed \cite{mansouri77}. These test theories are considered useful to examine potential alternatives to SRT. On the other hand, the reconstruction of the SRT, on the basis of the Lorentz-Poincar$\acute{e}$ scheme implies in an undetectable ether rest frame (non ether drift) at least in the first order \cite{lorentz}.

This comportment of the Lorentz-Poincar$\acute{e}$, as well as, of the Einstein theories arise because they do not rule the whole physics, for instance they do not involve gravitation.
It is also well known that the presence of a gravitational field breaks the Lorentz symmetry. 
For instance, the Sagnac effect \cite{sagnac13} shows that two opposite light beams travel in different time intervals the same closed path on a rotating disk. 
This effect is difficult to reconcile with SRT, because on the basis of the Lorentz transformation an isotropic light propagation is required in each point of the rotating disk, and a null effect is predicted. It is necessary to invoke the equivalence principle between an accelerated system and a gravitational field in order to take into account this effect. However, in the generalized Sagnac effect \cite{wang04} observed in a light waveguide loop consisting of linearly and circularly segments, any segment of the loop contributes to the phase difference between two counter-propagating light beams in the loop, and the situation is again difficult to reconcile with relativistic theories. On the other hand, according to Fox \cite{fox64},it is possible to preserve the general Lorentz Poincar$\acute{e}$ symmetry group without assuming the constancy
of light speed. There are also the so called extended theories, where the SRT is modified in order to including the Planck scale parameters \cite{camelia01}(double relativity theories), suggesting several dispersion relations that include theories where an energy dependent speed of light \cite{magueijo99} is claimed.

There is also evidences suggesting that the propagation of light over cosmological distances has anisotropic characteristics\cite{nodland97}, with dependence on direction and polarization. This picture is in agreement with the interpretations of the COBE\cite{smoot91} measurements giving the Earth's ``absolute'' velocity in relation to the uniform cosmic microwave background radiation(CMBR). Of course, there are also interpretations claiming that the COBE measurements give only a velocity for the ``relative'' motion between the Earth and the CMBR \cite{yaes93}. For instance, it is possible to remove the Earth motion to obtain a ``virtual'' image, where an isotropic distribution of CMBR with small fluctuations ($\delta T/T \sim 10^{-5}$) can be seen.
There is also some indirect evidences \cite{hatch02} via analysis of the Global Positioning System (GPS)
for the presence of an ether-drift velocity. 
The assumption of a preferred frame agrees also with a previous analysis made by Brans and Stewart \cite{brans73} where a description of the topology of the universe has imposed a preferred state of rest so that the principle of special relativity, although locally valid, is not globally applicable. 

So far, several tests about violation of the isotropy of the speed of light have been made. In most cases, the tests involve the so called round-trip test of light-speed isotropy like Michelson-Morley experiment and all its variants. Particularly, Miller \cite{miller33,cahill05} has claimed a non-null results in the M-M experiments. These aspects are presented in Appendix A.   

On the other hand, there are also several one-way test of light isotropy experiments. In most cases, they have claimed
a null result \cite{cialdea72, krisher90, turner64, gagnon88}, and some have claimed success \cite{kolen82,silvertooth86}. Particularly, Silvertooth has claimed an experimental detection of the ether drift velocity using a device capable of detecting the beams arriving in opposite directions\cite{silvertooth86}. Silvertooth report in 1986 a light anisotropy toward the direction of Leo constellation and compatible with COBE results. The experiment is an unusual double interferometer, an arrangement of light paths and detectors hard to be reproduced. In addition, the presence of a feedback into the laser is quite probable.

In this paper, we report results of a search for anisotropic light propagation as a function of the laser beam alignment relative to the Earth's velocity vector, using a diffraction device. The method is based on simple optical principles. Initial attempts have used digital images of the diffraction spots. However, this method was working in the limit of sensitivity. In other words, the signal's size was close to the measurement resolution. Now, our results are obtained by using the highly sensitive segmented photo-diodes to measure the position of diffracted light spots. In Section 2, the experimental setup and the basic operating principles of the diffractometer are presented. The Earth's velocity vector on the basis of the Doppler shift of the CMBR results are presents in Section 3. In Section 4, the two scanning methods and their results are presented, and finally Section 5 contains our conclusions.  

\section{Experimental setup and method}

The diffraction experiment is installed on the campus of the Universidade Federal
Fluminense, Niter\'oi, Rio de Janeiro-Brazil at sea level. The position is: latitude:
$22^{0}54^{\prime}33^{\prime\prime}$ S, longitude: $43^{0}08^{\prime}39^{\prime\prime}$ W. The diffraction experiment is mounted on a horizontal rotating circular table.
The layout of the diffraction device is shown in Fig.1. A laser beam transverse to a diffraction grating is diffracted
in several rays. In order to determine the position of the light spots, we have used two segmented PSD photo-diodes divided into two segments, separated by a gap (see Fig.2). 
The position of each photo-diode coincides with the positions of the maxima intensity of the diffraction images,
for $n=+1$ and $n=-1$ respectively, as shown in Fig.1. 
Two precision multi-axis positioning systems, and each one consist of a vertical platform with two independent (X-Y) micrometers, have been used to mount the photo-diodes.  

Following Fig.1, it is possible to see that the maxima of intensity of the diffraction images (rays) satisfy the condition
\begin{equation}
\sin \alpha_n=\pm n\frac{\lambda}{\eta\;\delta},\;\;\;with\;\;(n=0,1,2,....),
\end{equation}
where $\lambda(=632.8\,nm)$ is the wave length, $\delta(=1/600\,mm)$ is the diffraction grating step
and $\eta(=1.000226)$ is the refraction index of air. 
The wave length $\lambda$ can be obtained as the ratio between the speed of light $c$ and the light frequency
$\nu$ resulting in $\lambda=c/\nu$. An expression for $c$ as a function
of the angle $\alpha$ can be obtained as 
\begin{equation}
c=\frac{\eta \nu \delta}{n}\sin(\alpha_n).
\end{equation}
Under the assumption that $\nu$ and $\eta$ remain constant during the experiment, and if c depends on the direction of propagation, variations of the diffraction spot positions, $\alpha_n$, 
for instance, as a function of the laser beam alignment, relative to the Earth's velocity vector, can be interpreted as 
an indication of violation of the isotropy of $c$. The relative variation can be expressed as
\begin{equation}
\frac{\Delta c}{c}=\frac{\Delta (\sin \alpha_n)}{\sin \alpha_n}=\cot \alpha_n \;\Delta \alpha_n,
\end{equation}
We look for this anisotropic light propagation signal through measurements of $\Delta \alpha_n$ as a function of the Earth's velocity vector. The search has been made by using two independent types of scanning and the methods, as well as, the results are presented in section 4.

The determination of the position of the light spots is made by measuring the output photo-current in each segment of the photo-diodes. A symmetric spot, positioned at the center, generates equal photo-currents in the two segments. The relative position is obtained by simply measuring the output current of each segment. The position of a light spot with respect to the center on a segmented photo-diode is found by 
\begin{equation}
\Delta l=l_0\left(\frac{I_1-I_2}{I_1+I_2}\right),
\end{equation}
where $l_0$ is a proportionality constant. The method offers position resolution better than $0.1 \mu m$, and the angular variation can be obtained as
\begin{equation}
\Delta \alpha_n = \frac{\Delta l}{R}=\frac{l_0}{R}\left(\frac{I_1-I_2}{I_1+I_2}\right).
\end{equation}
For the diffraction experiment with $R=30.0\;cm$, the angular resolution is better than $3.3\times 10^{-7}rad$.

We have used the data acquisition system of the Tupi muon telescope \cite{navia05,augusto05,naviab05}, which is made on the basis of the $Advantech$ $PCI-1711/73$ card. The analog output signal from each segmented photo-diodes is linked with the analog input of the $PCI$ card. The $PCI$ card has $16$ analog input channels with a $A/D$ conversion up to $100$ kHz sampling rate. All the steps as the addition and the subtraction of currents are made via software, using the virtual instrument technique. The application programs were written using the Lab-View tools. A summary of the basic circuit is shown in Fig.2.

\section{The Earth's velocity vector}

The discovery of a pervasive background radiation from the universe by Penzias and Wilson \cite{penzias65} in 1965 is probably
the strongest evidence for the hot Big Band model. The Cosmic Microwave Background Radiation (CMBR) is a $2.7$ Kelvin thermal black body spectrum with a peak in the micro wave range. 
The CMBR is considered a relic of the Big Bang. In the past when the Universe was much smaller, the radiation was also much hotter. As the Universe expanded, it cooled down until today. 

In Penzias-Wilson's data, the radiation appeared as highly isotropic.
However, in the next round of experiments \cite{conklin69,henry71,smoot77} temperature anisotropies were found. These anisotropies are expressed using the spherical harmonic expansion, and the Earth's motion with velocity $\beta=v/c$ relative to the CMBR rest frame of temperature $T_0$ produces a Doppler shift as
\begin{equation}
\frac{\Delta T}{T_0}=\beta \cos \theta+\frac{\beta^2}{2}\cos 2\theta+O(\beta^3)
\end{equation}
In Table 1, measurments of the velocity vector of the Earth (solar system) in several experiments in chronological order using the anisotropy of the CMBR are summarized.
Southern Hemisphere airborne measurements of the anisotropy in the cosmic microwave background radiation by Smoot and Lubin \cite{smoot79} (1979-Lima, Peru) are in essential agreement with previous measurements from the northern hemisphere and the first-order anisotropy is readily interpreted as resulting from a motion of the Sun relative to the background radiation. 

The COBE data \cite{smoot00} indicate a big temperature anisotropy in the cosmic background radiation shich is represented by a dipole form with an amplitude of $\Delta T/T_0=1.23\times 10^{-3}=0.123\%$. This arises from
the motion of the solar system barycenter, with a velocity $v=371\pm 0.5kms^{-1}$ ($\beta=0.001237\pm 0.000002$) at 68\%CL, relative to the so called ``CMBR rest frame'' and towards a point whose equatorial coordinates are
$(\alpha,\;\delta)=(11.20^h\pm0.01^h,\;-7.22^0\pm 0.08^0$). This direction points to the Leo constellation.
Recently, the WMAP \cite{bennett03} mission has improved the resolution of the angular power spectrum of the CMBR and has verified the COBE results.

\section{Raster search techniques}

We look for an anisotropy signal in the light propagation as a function of the Earth's velocity vector. At our 
latitude ($\sim 23^0 S$) there are two passages of the Leo constellation on the horizon at every 24 hours.  The first one is near the West direction and the second is $\sim 12$ hours later and it is near the East direction. Consequently it is possible to mount a laser diffraction experiment on a horizontal turntable system and point the laser beam toward the Leo constellation. The raster search can be made by using two methods as are described below.

\subsection{Passive raster search system due to Earth's rotation}

This method consists in to fix the laser beam direction toward the first or second passage of the Leo constellation
on the horizon. As the Earth rotates, the laser beam will be aligned to the first or second passage of the Leo constellation on the horizon over a 24 hour period. Thus, we have a raster search system due to the Earth's rotation including the COBE coordinates. Data collection usually begins at least 6 hours before the Leo constellation (COBE coordinates) culmination.
At the end of autumn in the southern hemisphere which is our case, we have a privilege 
first passage, because it happens in the first hours of the day (dawn) and the human activities in the laboratory and especially in the neighboring laboratories are at a minimum.

As the laser, the diffraction grating, and the detectors are always fixed, the method is free of mechanical perturbations, which can be introduced, for instance, when the system is rotated. However, the method requires measurements over a long period of time (at least 12 hours) and several days and this introduces the so called DRIFT-long-term timing variation by aging 
due to temperature variations (diurnal and semi-diurnal temperature dependence). In the case of diffraction experiments this effect is amplified due to the temperature dependence of the refraction index. 
There is also the JITTER-timing (short term) noise due to statistical fluctuations of the signal (shot and thermal noises) and they have a ``white'' frequency distribution, i.e. the spectral power densities are constants.

Examples of such raster scans during the first passage of the Leo-CC (Leo-COBE coordinates) on the horizon are shown from Fig.3 through Fig.5. The raw data are obtained by reading the signal output at every ten seconds (see upper picture in Fig.3), and the histograms are obtained by averaging the raw data in blocks of 10 minutes. In all cases, a peculiar signature is observed, the Leo-CC culmination on the horizon is always between a peak and a depression. 
If the COBE-CC has an altitude, h, and an azimuth angle, $\theta_A$, the projection of the Earth's velocity, $v$, on the laser beam direction is
\begin{equation}
v_p=v \cos(h)\cos(\theta_A-\theta_{beam})
\end{equation}
where $\theta_{beam}$ is the azimuth of the laser beam (Leo-CC azimuth on the horizon). It is possible to see that the sensitivity of the diffraction experiment to the Earth's velocity for this scan system 
begins when the altitude on the horizon of the Leo-CC is $h_1 \sim 20^0$ and it finishes when the altitude is $h_2 \sim -20^0$. In this passage on the horizon, the Leo-CC has a variation in its azimuthal angle of around $17^0$, and it is equivalent to an azimuthal rotation of the diffractometer (around $17^0$). The azimuthal dependences of the two passages of COBE-CC on the horizon on 2006/06/16 are shown schematically in the upper 
drawing of Fig.6, and the details of the first passage are shown in the lower drawing.
In $h_1$ the $\cos \theta_A=-0.035$ and in $h_2$ the $\cos \theta_A=-0.317$ (smaller than 1 order of magnitude), and it is responsible for this peculiar structure with a peak followed by a depression. 

The effect provides a dipole behavior ($\propto \cos\theta_A$) observed in a narrow angular window of ($\sim 17^0$) centered in the azimuth angle of the Leo-CC.
We have obtained also some examples of the raster scan during the second passage of the Leo-CC on the horizon, as shown in Fig.7. As already expected, there is a depression followed by a peak as in Fig.6.
In order to extract the Earth's velocity from these experimental data, it is necessary to remove the DRIFT-long-term timing variation, because they are obtained in different days. Meantime this procedure is not free from experimental bias. The calibration will be done in the next search with active rotation which is free of the DRIFT-long-term timing variation.    
   
\subsection{Active raster search system with rotating turntable}

In this active system, the laser beam is first pointed toward the direction of the Leo constellation (Leo-CC) when it is exactly on the horizon. Then the turntable, upon which the entire laser diffraction experiment is mounted, is rotated in steps of 30 degrees up to 180 degrees. At every angular step, the output current of the photo-diodes is registered during one minute at a counting rate of 10 readings per second. A complete set of measurements can be done in less than ten minutes. Consequently the measurements are free from
 DRIFT-long-term timing variations. They are influenced only by the JITTER-timing uncertainties (noise in the system by statistical fluctuations of the signals). 
However, this method requires a careful rotation of the system in order to avoid mechanical perturbations.
The measurements, after a gaussian fit in the raw data, are shown in Fig.8 for seven angular regions, and
the one-way light path anisotropy can be extracted from
\begin{equation}
\frac{\Delta c}{c}=\cot \alpha \Delta \alpha,
\end{equation}
with
\begin{equation}
\Delta \alpha=\frac{l_0}{R}\left[\frac{I_1-I_2}{I_1+I_2}\right],
\end{equation}
where the calibration factor obtained for these measurements is $l_0(=0.407\pm 0.034mm)$. 

We have analyzed the experimental data using the test theory of Mausouri and Sexl\cite{mansouri77} where the one way speed of light is anisotropic by the among
\begin{equation}
c(\theta)=c-v(1+2a)cos \theta,
\end{equation}
where $\theta$ is the angle between the velocity, $v$, of the moving system (i.e. Earth motion) and the direction of light propagation. The value $a=-1/2$ corresponds to the isotropic SRT prediction, and $a\neq -1/2$ represents an anisotropic signal in the one-way path speed of light. The parameter $a$ can be obtained by fitting the test theory to the experimental results using the expression
\begin{equation}
\frac{\Delta c}{c}=\cot \alpha \Delta \alpha=-\beta (1+2a)cos\theta,
\end{equation} 
where $\beta(=0.001237\pm 0.000002)$ is the COBE Earth's velocity parameter. The comparison between our measurements and the test theory is shown in Fig.9, where an offset such that $(I_1-I_2)/(I_1+I_2)=0$ at $\theta=90^0$ has been used. The experimental results seem to agree to a $\beta (1+2a)\cos \theta_A$ signature, and the parameter $a$ extracted from our data is 
\begin{equation}
a=-0.4106\pm 0.0225,
\end{equation}
which differs from the $a=-1/2$ SRT prediction, as well as, from some experimental upper limits using the Mossbauer
effect \cite{turner64}. 

\section{Conclusions and remarks}

The discovery of a dipole anisotropy in the CMBR is interpreted as a Doppler shift produced by the Earth' motion (solar barycenter). An experimental survey has been made in order to test if the Earth's velocity is relevant on light propagation in a quantity of first order. The measurements have been obtained by using a laser diffraction experiment mounted on turntable system. Two optic sensors (segmented photo-diodes) were
used for measuring the position of diffracted light spots with a precision better than  $0.1\;\mu m$. 
The experiment is easy to reproduce, and it is based on simple optical principles. Two raster search techniques have been used to look for signals of anisotropic light propagation as a function of the laser beam alignment relative to the Earth's motion. The results obtained with both methods show that the course of the rays is affected by the motion of the Earth. 

In the scan subjected to Earth's rotation, the sensitivity of the diffraction experiment to the Earth's velocity 
begins when the altitude on the horizon of the Leo-CC is $h_1 \sim 20^0$, and it finishes when the altitude is $h_2 \sim -20^0$. In this passage on the horizon, the Leo-CC has a variation in its azimuthal angle of around $17^0$, and it is equivalent to an azimuthal rotation of the diffractometer by the same angle (around $17^0$).
This effect provides a dipole behavior ($\propto \cos \theta_A$) in a narrow angular window ($\Delta \theta_A \sim 17^0$), and it is observed as a peak followed by a depression in the first passage, or a depression followed by a peak in the second passage (see Fig.6 upper). 

In the scan subjected to an active rotation, from zero to 180 degree relative to the Earth's velocity direction, a complete set of measurements can be done in $10$ minutes. Consequently it is free from DRIFT-long-term timing variations.
The experimental data are susceptible of being interpreted by the test theory of Mausouri and Sexl\cite{mansouri77}
where the one way speed of light is anisotropic by the amount $c(\theta)=c-v(1+2a)cos \theta$. The parameter $a$ extract from a fit of the data is $a=-0.4106\pm 0.0225$, 
and it differs from SRT prediction with $a=-0.5$ by an amount of 18\%.

We remark that the CMBR dipole is a frame dependent quantity. We can thus determine the ``absolute rest frame'' of the universe as that in which the CMBR dipole would be zero. In short, our results point out that it is not possible to neglect the preferred frame imposed by  cosmology.

\section{Acknowledgments}

This paper is a memorial tribute to our professor and friend C.M.G. Lattes, who introduced us the non-conventional aspects of relativity. We are thankful to all the members of the laboratory of Thin Films of IF-UFF for implementing the techniques of using segmented photo-diodes.

\appendix
\section{The Miller's ether drift direction}

According to the theories that incorporate the length contraction principle (Einstein and Lorentz-Poincar$\acute{e}$ theories), experiments where two orthogonal light paths are compared (two way speed experiments) like the Michelson-Morley interferometer and all variants are incapable of detecting the Earth's motion (no ether drift) due to the length contraction of the interferometer arm parallel to the direction of the Earth's velocity.

Strictly speaking, a null result is expected only in vacuum where the refractive index is $\eta=1$.
While, if $\eta\neq 1$ the Fresnel's drag effect in the rest frame of the medium ($\Sigma$) cancel the effect of the genuine Lorentz transformation to a moving frame ($\Sigma'$).
Following the Lorentz transformation equations from $\Sigma'$ with speed $v$ to $\Sigma$, and taking into account the Fresnel relation of the speed of light in the medium $c'=c/\eta$ it is possible to obtain the so called two-way speed of light anisotropy as
\begin{equation}
\bar{c'}(\theta)=\frac{2c'(\theta)c'(\theta+\pi)}{c'(\theta)+c'(\theta+\pi)},
\end{equation}
and the relative variation as
\begin{equation}
\frac{\Delta c'(\theta)}{c}=-\frac{v^2}{c^2}\left[\frac{\eta^2-1}{\eta^2}(1-\frac{3}{2}\sin \theta^2)\right].
\end{equation}
We can see that the two-way speed of light anisotropy is null only in the case $\eta=1$ (vacuum). This prediction is in agreement
with modern ether drift experiments in vacuum \cite{brillet79, muller03}, using two cavity-stabilized lasers and whose value is
\begin{equation}
\frac{\Delta c'}{c} \sim 10^{-15}.
\end{equation}
In the gaseous mode, for instance air ($\eta=1.000226$), a maximum value of $\Delta c'/c$ 
happens in reference axis parallel to Earth's velocity.
 The tine fringe shifts observed in various Michelson-Morley type experiments, represent a non-null effect for the two-way speed of light anisotropy. Dayton Miller\cite{miller33,cahill05} was one of the first few in claiming that the Michelson-Morley data and his own data obtained in the mount Wilson
are non-null results. Particularly, the mount Wilson data obtained in $1925-1926$ is compatible with an
observable Earth velocity of $v\sim 8.5 \pm 1.5 km/s$, when the data is analyzed on the basis of classical physics.
While on the basis of a different calibration including the length contraction (see eq.A2), the Miller result gives the Earth velocity $v=433 km\;s^{-1}$ in the direction whose equatorial coordinates are
$(\alpha = 5.2^h,\;\delta = -67^0)$ which coincide with the direction of the Dorado constellation.

Now we know that Miller's determination of the velocity direction of the Earth is incompatible with modern observations. The Miller's direction for the Earth velocity is almost perpendicular to the direction established by COBE, observing the CMBR anisotropy. A second aspect against Miller's result is that a re-analysis of the Miller's data by Shankland et. al. \cite{shankland55} indicates that Miller's result must be invalid because the temperature variations were not taken into account correctly. 
In our opinion, Miller's result has the same problem as the first results of the CMBR survey as is shown in Table 1. For instance, both Miller and Conklin have obtained a non-null result on the two-way path light speed anisotropy and the dipole anisotropy of the CMBR, respectively. Nevertheless, both experiments have failed to obtain the coordinates of the Earth's velocity vector direction correctly.

\newpage
\begin{table}
\caption{Vector velocity of the Earth (solar system) relative to the CMBR rest frame, measured using the anisotropy of the CMBR in several experiments.}
\begin{ruledtabular}
\begin{tabular}{ccccc}
 %& \multicolumn{3}{c}{Muon telescopes at ground} \\
Observer & Year & $v_{E} (km\;s^{-1})$ & $\alpha$ (hour) & $\delta$ (degree)\\
\hline
Pensias $\&$ Wilson (ground)\cite{penzias65}     & 1965 & \multicolumn{2}{c} {Isotropic}  \\
\hline
Conklin (ground)\cite{conklin69}    &  1969 &  $200\pm 100$ & $13\pm 2$   & $30\pm 30$  \\
Henry (balloon)\cite{henry71}     &  1971 &  $320\pm 80 $ & $10\pm4$    & $-30\pm 25$ \\ 
Smoot et al. (airplane)\cite{smoot77}  & 1977 & $390\pm 60$ & $11.0\pm 0.5$ & $6 \pm 10$  \\
COBE (satellite)\cite{smoot00}  & 1991 & $371\pm 0.5$ & $11.20\pm 0.01$ & $-7.22\pm 0.08$ \\
WMAP (satellite)\cite{bennett03}  & 2003 & $368\pm 0.2$ & $11.20\pm 0.01$ & $-7.22\pm 0.08$ \\
\end{tabular}
\end{ruledtabular}
\end{table}

\newpage
%%%%%%%%%%%%%%%%%%%%%%%%%%%%%%%%%%%%%%%%%%%%%%%%%%%%%%%%%%%%%%%%%%%%%%%%%%%%%%%%%%%%%%%% 
\vspace*{+8.0cm} 

\begin{figure}[th]
\vspace*{-4.0cm}
\includegraphics[clip,width=0.8
\textwidth,height=0.8\textheight,angle=0.] {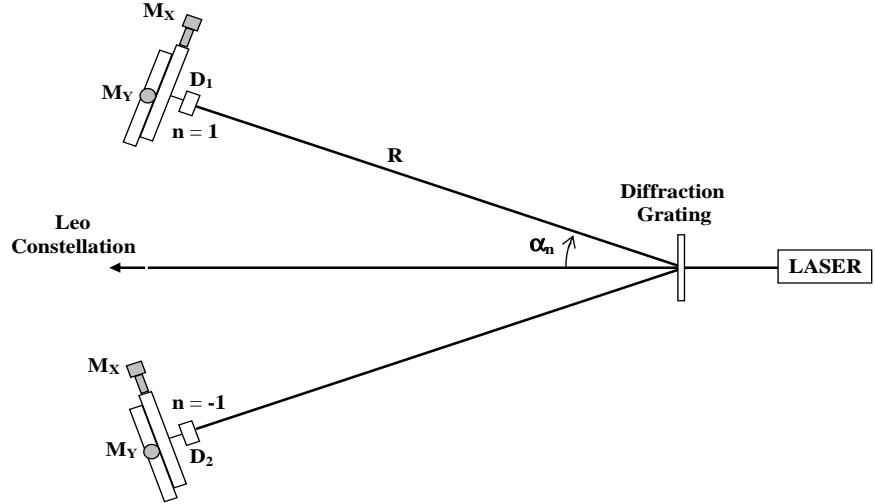}
\vspace*{-6.0cm}
\caption{General layout of the diffraction experiment. Two segmented photo-diodes ($D_1$ and $D_2$) are positioned using two vertical platforms with two  positioning system (micrometers $M_x-M_y$) to detect two diffracted rays produced by a HE-Ne laser on a grating diffraction device. The relative position of a light spot with respect to the center on a segmented photo-diode is obtained by simply measuring the output current of each segment. The setup is mounted on a turntable system}
\end{figure}

\begin{figure}[th]
\vspace*{-4.0cm}
\includegraphics[clip,width=0.8
\textwidth,height=0.8\textheight,angle=0.] {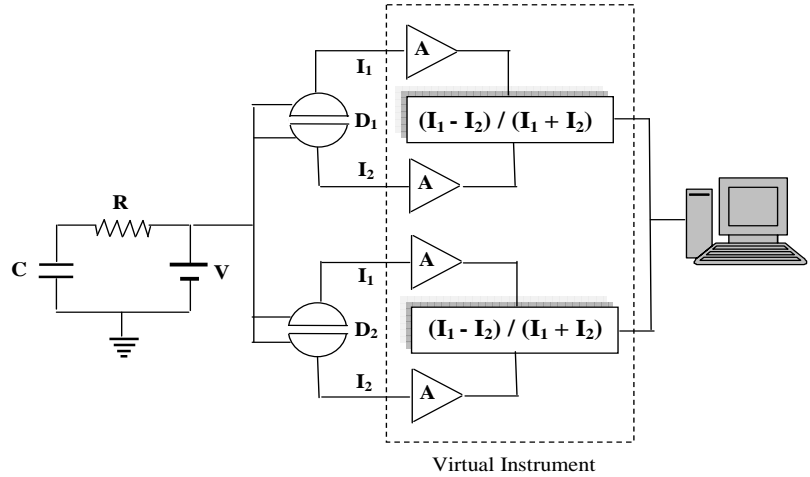}
\vspace*{-6.0cm}
\caption{Block diagram of the diffraction experiment data acquisition system. $D_1$ and $D_2$ represent the segment photo-diodes.}%
\end{figure}

\begin{figure}[th]
\vspace*{-0.0cm}
\includegraphics[clip,width=0.7
\textwidth,height=0.7\textheight,angle=0.] {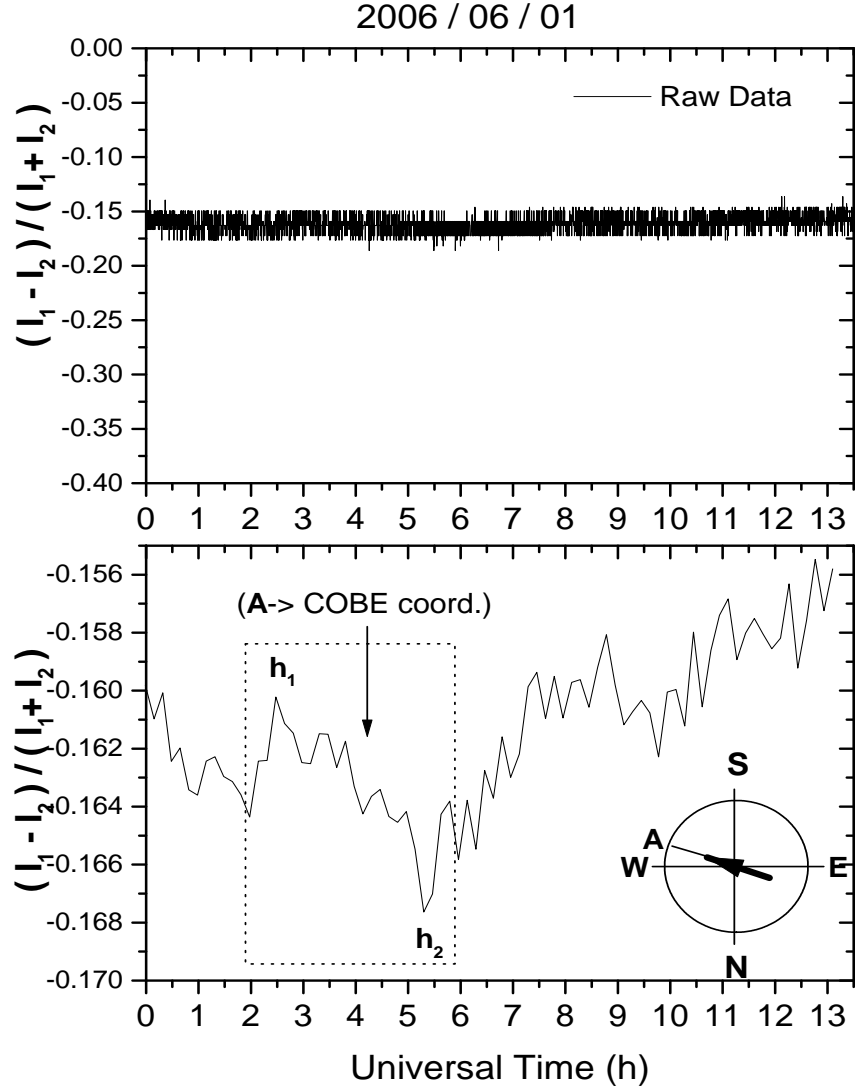}
\vspace*{-1.0cm}
\caption{Upper figure: Raw data obtained as a time series by measuring the photo-diodes currents at every 10 s (counting rate of $0.1Hz$).
Lower figure: Histogram obtained averaging the raw data in blocks of 10 minutes.
The vertical arrow indicates the moment of the passage of the Leo constellation for the horizon, and the arrow inside of the circle represents the orientation of the laser beam. $h_1\sim 20^0$ and $h_2\sim -20^0$ are the Leo altitude before and after the Leo-CC culmination on the horizon.}%
\end{figure}

\begin{figure}[th]
\vspace*{-1.0cm}
\includegraphics[clip,width=0.7
\textwidth,height=0.7\textheight,angle=0.] {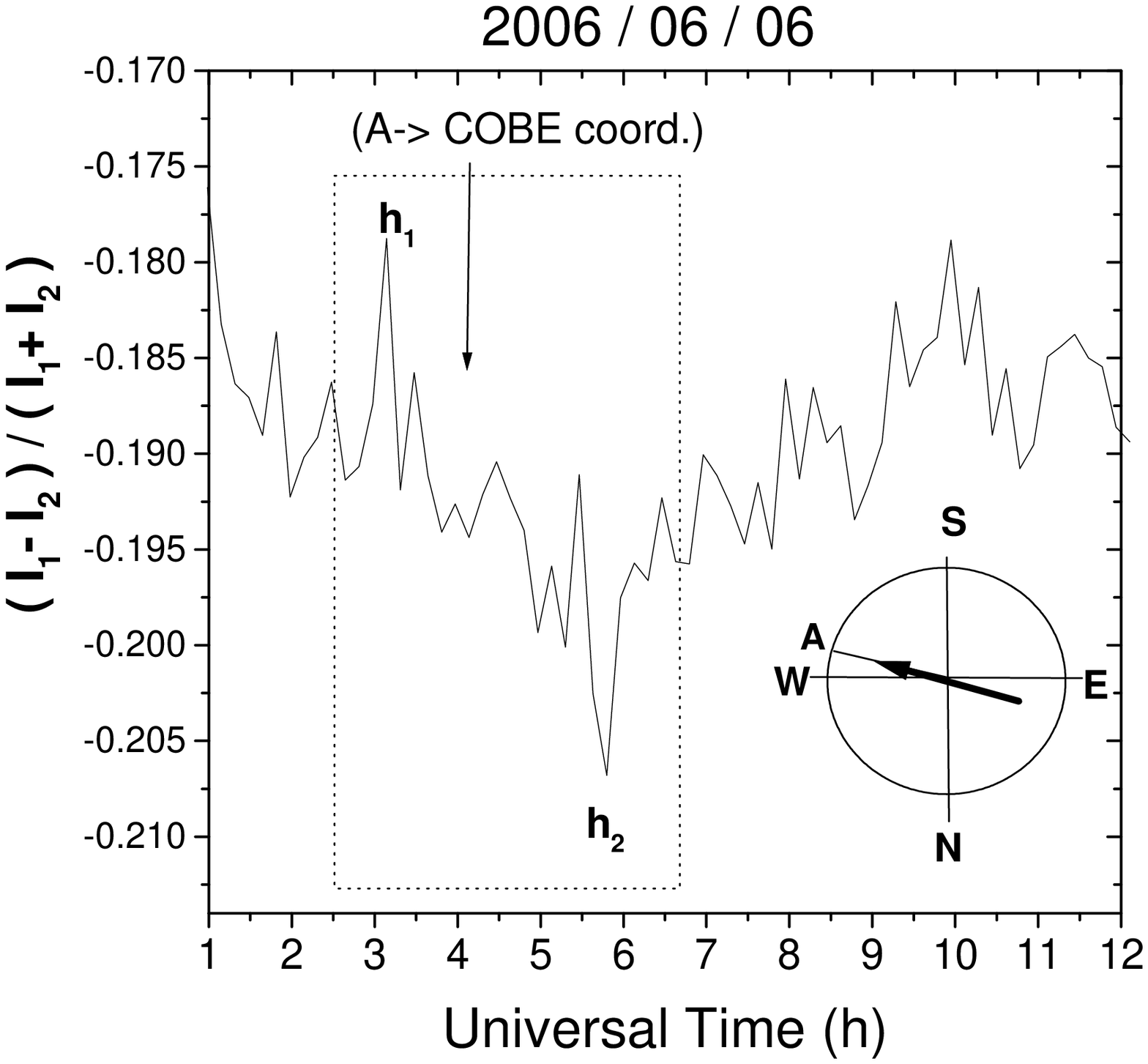}
\vspace*{-6.0cm}
\caption{Histogram obtained averaging the raw data in blocks of 10 minutes.
The vertical arrow indicates the moment of the passage of the Leo constellation for the horizon, and the arrow inside of the circle represents the orientation of the laser beam. $h_1\sim 20^0$ and $h_2\sim -20^0$ are the Leo altitude before and after the Leo-CC culmination on the horizon.}%
\end{figure}

\begin{figure}[th]
\vspace*{-3.0cm}
\includegraphics[clip,width=0.6
\textwidth,height=0.6\textheight,angle=0.] {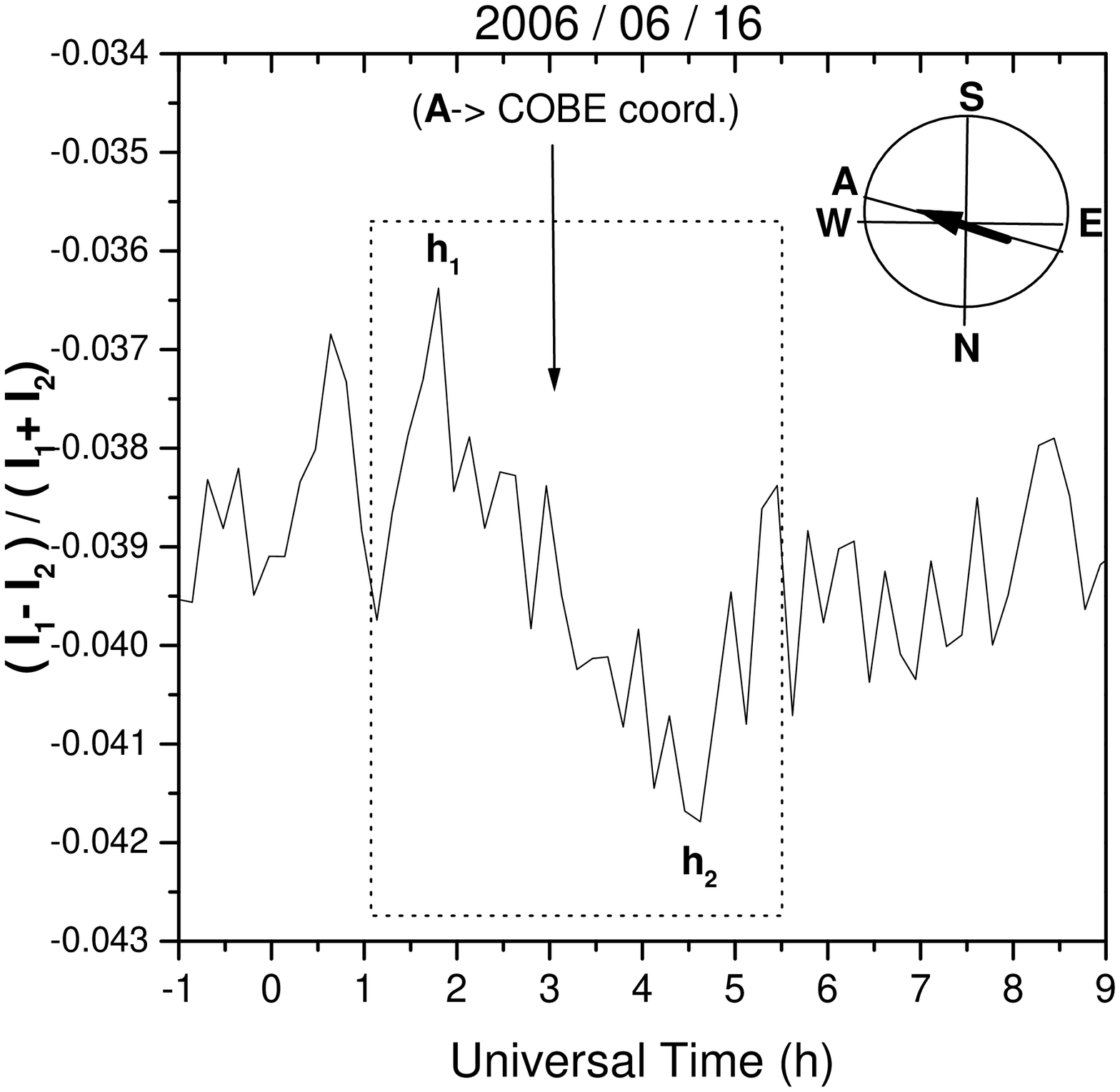}
\vspace*{-4.0cm}
\caption{Histogram obtained averaging the raw data in blocks of 10 minutes.
The vertical arrow indicates the moment of the passage of the Leo constellation for the horizon, and the arrow inside of the circle represents the orientation of the laser beam. $h_1\sim 20^0$ and $h_2\sim -20^0$ are the Leo altitude before and after the Leo-CC culmination on the horizon.}%
\end{figure}

\begin{figure}[th]
\vspace*{-0.0cm}
\includegraphics[clip,width=0.7
\textwidth,height=0.7\textheight,angle=0.] {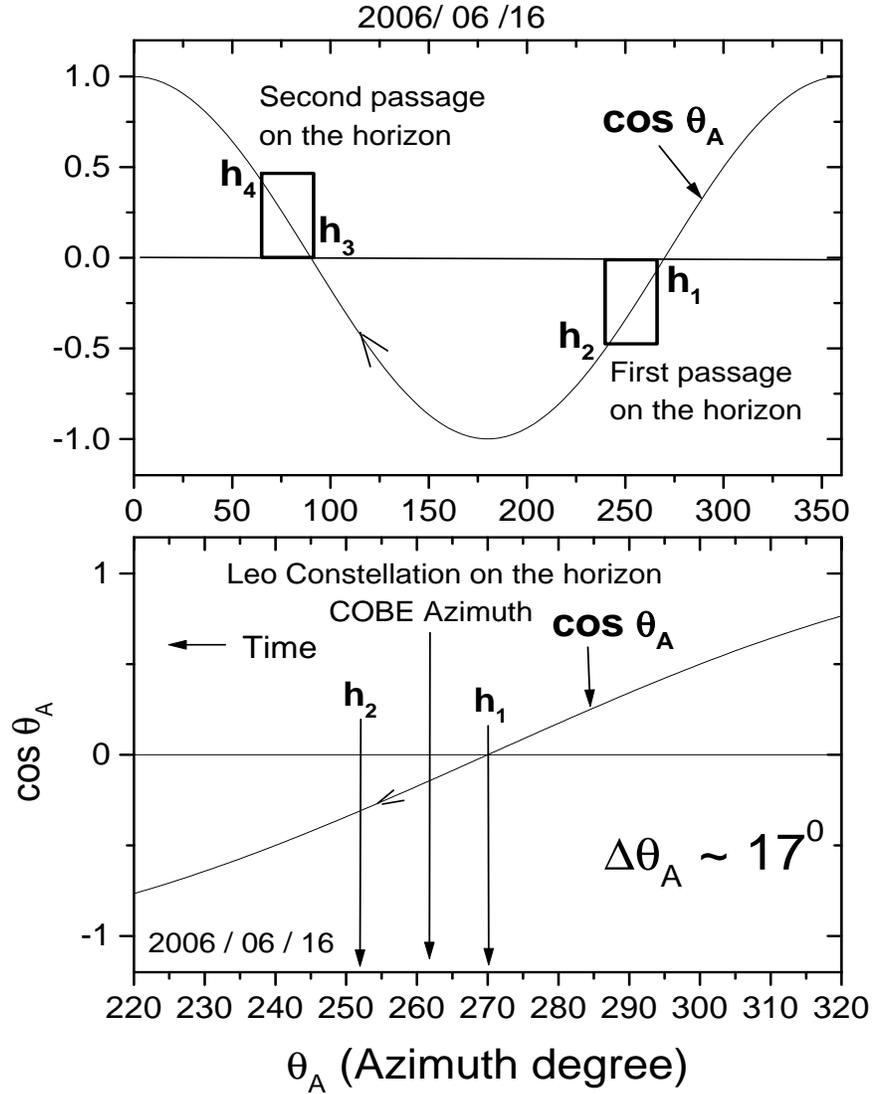}
\vspace*{-1.0cm}
\caption{Upper: Azimuthal variation of the Leo constellation (COBE coord.) during the two passages on the horizon on $2006/06/16$ and at Rio de Janeiro latitude. The two
azimuthal narrow windows are indicated by the two squares between the Leo altitudes $h_1$ to $h_2$ and $h_3$ to $h_4$ and represents the regions where the diffraction experiment is sensitive to the Earth's velocity. Lower: 
Here is a zoom of the first passage in the upper figure.}%
\end{figure}

\begin{figure}[th]
\vspace*{-0.0cm}
\includegraphics[clip,width=0.7
\textwidth,height=0.7\textheight,angle=0.] {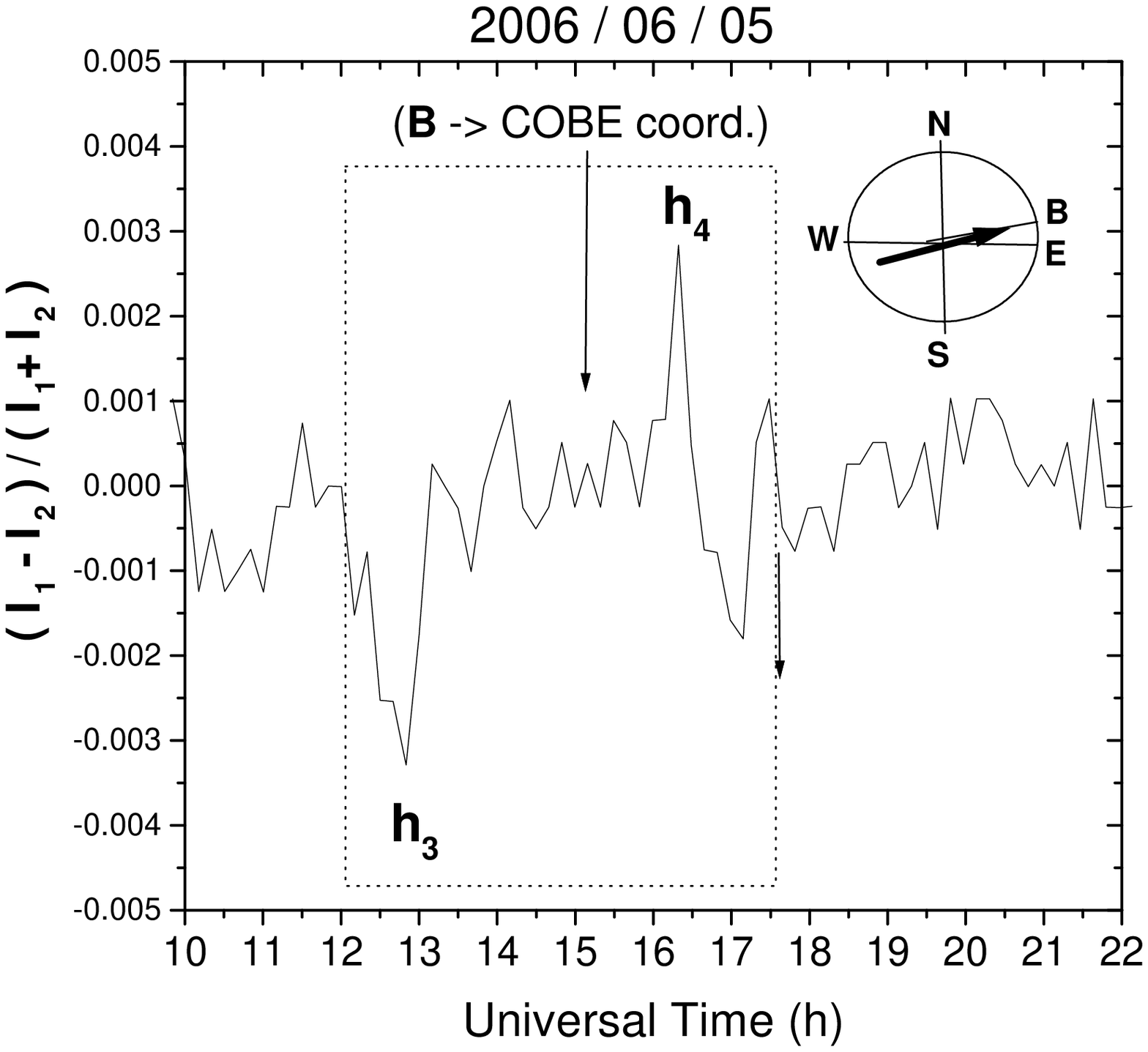}
\vspace*{-5.0cm}
\caption{Histogram obtained averaging the raw data in blocks of 10 minutes (during the second passage).
The vertical arrow indicates the moment of the passage of the Leo constellation for the horizon, and the arrow inside of the circle represents the orientation of the laser beam. $h_3\sim -20^0$ and $h_4\sim 20^0$ are the Leo altitude before and after the Leo-CC culmination on the horizon.}%
\end{figure}

\begin{figure}[th]
\vspace*{-0.0cm}
\includegraphics[clip,width=0.7
\textwidth,height=0.7\textheight,angle=0.] {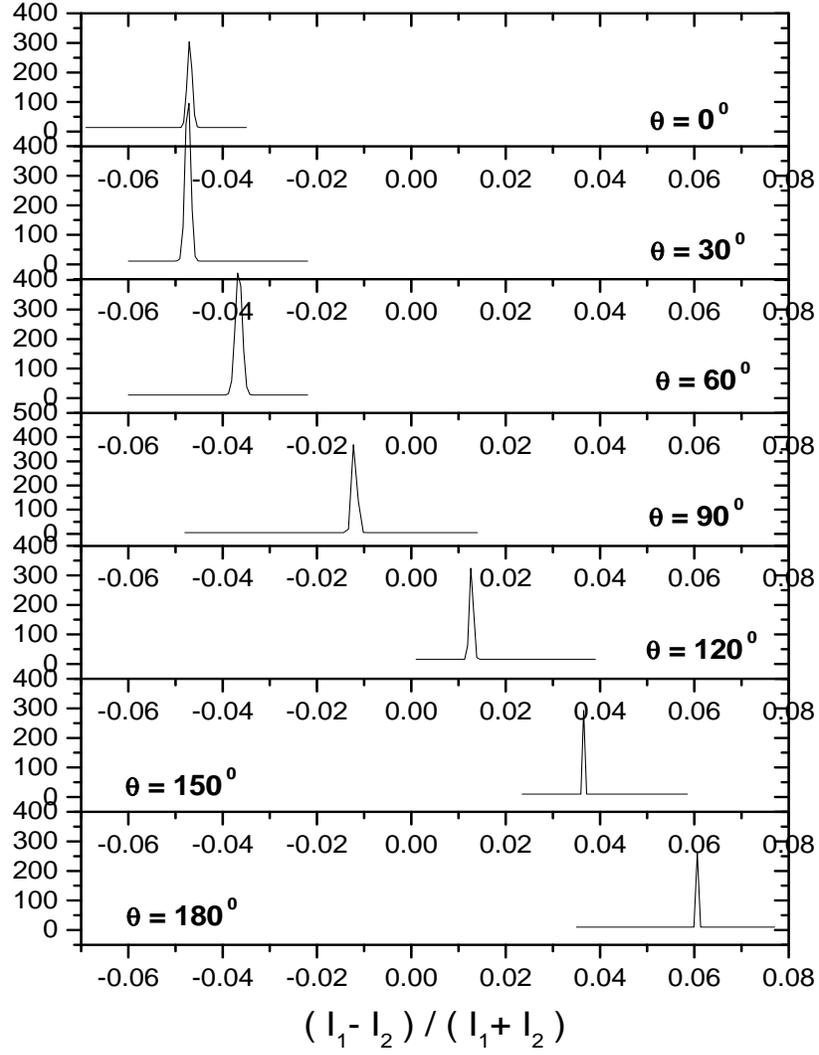}
\vspace*{-1.0cm}
\caption{Counting rate plotted as a function of $(I_1-I_2)/(I_1+I_2)$ at a given laser beam alignment relative to the Earth's velocity vector. The complete set of measurements was made in ten minutes.}%
\end{figure}
\begin{figure}[th]
\vspace*{-0.0cm}
\includegraphics[clip,width=0.7
\textwidth,height=0.7\textheight,angle=0.] {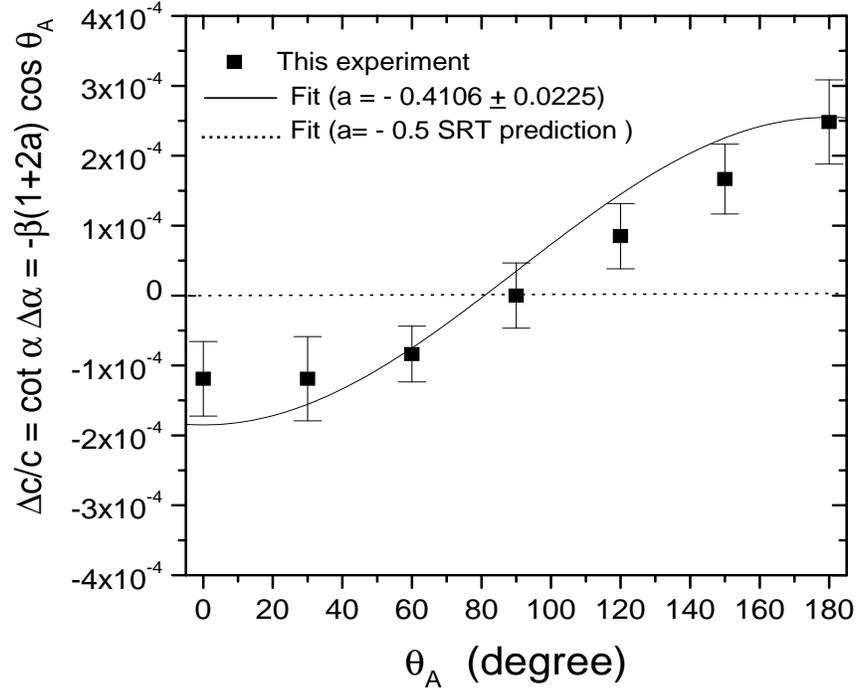}
\vspace*{-4.0cm}
\caption{Comparison between the one-way path light anisotropy  $\Delta c/c=-\beta (1+2a)\cos \theta$ function,  relative to the Earth's velocity vector and the experimental data, for two different values  
of the fit parameter, $a=-0.4106$ and $a=-1/2$ respectively. Here $\theta_A=0$ represent the laser beam pointed to the Leo constellation (Leo-CC).}%
\end{figure}

\end{document}